\documentclass[preprint,12pt]{aastex}
\usepackage{graphicx}
\usepackage{natbib}
\usepackage{lscape}

\shorttitle{ROTATION OF JETS FROM YOUNG STARS}
\shortauthors{COFFEY ET AL 2003}

\begin{document}

\title{Rotation of Jets from Young Stars: \\ New Clues from the Hubble Space Telescope 
Imaging 
Spectrograph \footnote{Based on observations made with the NASA/ESA $\it{Hubble}$ 
$\it{Space}$ $\it{Telescope}$, 
obtained at the Space Telescope Science Institute, which is operated by the Association of 
Universities for Research in 
Astronomy, Inc., under NASA contract NAS5-26555.}}

\author{
Deirdre Coffey\altaffilmark{1}, Francesca Bacciotti\altaffilmark{2}, Jens 
Woitas\altaffilmark{3}, Thomas 
P. 
Ray\altaffilmark{1}, Jochen Eisl\"{o}ffel\altaffilmark{3}}

\altaffiltext{1}{Dublin Institute for Advanced Studies, 5 Merrion Square, Dublin 2, 
Ireland \email{dac, 
tr@cp.dias.ie}}
\altaffiltext{2}{I.N.A.F. - Osservatorio Astrofisico di Arcetri, Largo E. Fermi 5, 50125 
Firenze, Italy 
\email{fran@arcetri.astro.it}}
\altaffiltext{3}{Th\"{u}ringer Landessternwarte Tautenburg, Sternwarte 5, 07778 
Tautenburg, Germany 
\email{woitas, 
jochen@tls-tautenburg.de}}

\begin{abstract}
We report findings from the first set of data 
in a current survey to establish conclusively whether jets from young stars rotate. We 
observed 
the bi-polar jets from the T Tauri stars TH28 and RW Aur,  and the blue-shifted jet from T 
Tauri star 
LkH$\alpha$321, 
using 
the Hubble Space Telescope Imaging Spectrograph (HST/STIS). 
Forbidden emission lines (FELs) show distinct and systematic velocity asymmetries 
of 10 -- 25 ($\pm$ 5) km~s$^{-1}$ at a distance of 0$\arcsec$.3 from the source, 
representing a
(projected) distance of $\approx$ 40 AU along the jet 
in the case of RW Aur, $\approx$ 50 AU for  TH28, and 
165 AU in the case of LkH$\alpha$321. 
These velocity asymmetries are interpreted as 
rotation in the initial portion of the jet where it is accelerated and collimated. 
For the bi-polar jets, both lobes appear to rotate in the same direction. Values obtained 
were in 
agreement with the 
predictions of MHD disk-wind models \cite{Bacciotti02, Anderson03, Dougados03, Pesenti03}. 
Finally, we determine, from derived toroidal and poloidal velocities, values for the 
distance from the 
central axis 
of the footpoint for the jet's low velocity component of $\approx$ 0.5 - 2 AU, consistent 
with the models 
of 
magneto-centrifugal launching \cite{Anderson03}.  \end{abstract}

\keywords{ISM: jets and outflows --- stars: formation, pre-main sequence --- stars: 
individual: TH28 --- stars: individual: RW Aur --- stars: individual: LkH$\alpha$321}


\section{Introduction}

A key question in star formation research concerns the mechanisms behind the 
launch of jets from young 
stars.  These jets are believed to play an important role in the removal of excess angular 
momentum from 
the 
system, thus allowing  accretion of matter onto the star up to its final mass. 
It is generally acknowledged that magneto-centrifugal forces are responsible for jet 
launching. In 
particular in the 
so-called 
`disk-wind' model (e.g. \citealp{Ferreira97}; \citealp{Konigl00}) 
the jet is launched 
from the disk surface within a few AU from the star; while in the 'X-wind' model 
\cite{Shu00} the base 
of the flow is 
located at a few stellar radii from the source.   
To date, resolution constraints on observations have 
impeded progress in validating the magneto-centrifugal mechanism, since jet 
launching occurs on small scales (i.e. less than 20 AU from the star); moreover, 
infall and outflow kinematics are 
complex and confused close to the source, that is often heavily embedded. 
Recently, however, interesting results have been obtained 
from observations of jets from more evolved, less embedded 
T Tauri stars (TTSs) for which the jet can be optically traced 
back to its origin.

Observational backing for canonical models would require, for example, proof of 
rotation around the symmetry axis, close to the base where the jet is launched. 
The first hint of jet rotation was reported for the HH 212 system \cite{Davis00}. 
However the knots were located 
at 2$\times$10$^3$ - 10$^4$ AU from the jet source. 
Independently,  asymmetries in velocities within the 
first 110 AU of the outflow 
from the T Tauri star DG Tau were found \citep{Bacciotti02}, indicative of rotation.
These results were obtained through an analysis 
of high angular resolution spectra taken 
with the Hubble Space Telescope Imaging Spectrograph (HST/STIS), aimed at 
probing the acceleration and collimation region of a stellar jet.  
Further confirmation of the rotation hypothesis came from Owens
Valley Radio Observatory (OVRO) observations, which report the sense of rotation of the 
disk of DG Tau 
to be the same as that of the jet \cite{Testi02}. Moreover, the derived toroidal 
velocities in the 
observed
portion of the jet were seen to be in agreement with the predictions of the 
magneto-centrifugal
models, and indeed they can be used to find the location on the disk plane of
the launching point of the wind (\citealp{Bacciotti02, Anderson03, Dougados03, 
Pesenti03}). 

These results motivated us to conduct an optical STIS survey to establish 
conclusively whether jets from young stars rotate. 
We report findings for another three sources (of eight in the survey sample) 
for which the data have already been acquired, i.e. the bi-polar jets from  the 
TTSs TH28 and RW Aur, and the blue-shifted jet from LkH$\alpha$321.

\begin{table*}
\begin{center}
\scriptsize{
\begin{tabular}{llllllllll}
\tableline \tableline
Star		&Location	&Distance	&Associated 	&M$_{star}$ 	&i$_{jet}$	
	&\.M$_{disk}$			&\.M$_{jet}$		&\.P$_{jet}$			
	&References\\ 
		&		&(pc)		&Outflow	&(M${_\odot}$)	&(deg)		
	&(M${_\odot}$yr$^{-1}$)		&(M${_\odot}$yr$^{-1}$)	&(M${_\odot}$ yr$^{-1}$ 
kms$^{-1}$)	\\ 
\tableline

TH28		&Lupus 3	&170 		&HH228		&...		& 10		
	&...				&3.4$\times$10$^{-8}$	&7.5$\times$10$^{-6}$   	
	&1, 2 \\

RW Aur		&Auriga 	&140 		&HH229		&$\sim$1		
&44			
&10$^{-6}$			&1.1$\times$10$^{-7}$	&3.8$\times$10$^{-5}$			
&3, 4 \\

LkH$\alpha$321	&Cygnus	 	&550 		&HH421		&...		&...			
&...		
		&...			&...					&5\\ 
\tableline
\end{tabular}
}
\end{center}
\caption{\scriptsize{
Sources studied in this paper. Where known, the table 
also lists the mass of the star, the jet inclination angle i$_{\rm jet}$ with respect to 
the plane of 
the sky, the mass accretion rate through the disk (\.M$_{disk}$), and the fluxes of mass 
(\.M$_{jet}$) and poloidal 
momentum (\.P$_{jet}$) in the jet. References: 1 - \citealp{BE99}; 2 - 
\citealp{Krautter86}; 
3 - \citealp{Woitas02}; 4 - \citealp{Martin03}; 5 - \citealp{Mundt98}.}
\label{list_sources}}
\end{table*}

\section{Observations}

Observations were made of the jets associated with TH28, LkH$\alpha$321 
and RW Aur using HST/STIS on June 22, August 20, and October 3, 2002 
respectively (proposal ID 9435). An aperture of 52 $\times$ 0.1 arcsec$^{2}$ 
was used with the G750M grating which gave a spectral sampling of 0.554 \AA
~pixel$^{-1}$, corresponding to a radial velocity of $\approx$ 25 km~s$^{-1}$ for 
the wavelength range covered.  The angular sampling was 0$\arcsec$.05 
pixel$^{-1}$. An acquisition of the stellar peak intensity prior to science 
observations allowed the slit to be centered accurately on the star before 
being offset to a position perpendicular to the jet axis at 0$\arcsec$.3 from 
the source. We assume inclination angles (w.r.t. the plane of the sky) of 10$\degr$ for 
TH28 
\cite{Krautter86}, 
44$\degr$ for RW Aur \cite{Martin03}, and 45$\degr$ for LkH$\alpha$321 (arbitrary, since 
unknown, 
although it may 
have a large inclination angle from spectro-astrometric measurements \cite{Whelan03}). The 
offset of 
0$\arcsec$.3 
then represents a deprojected distance of $\approx$ 51, 195 and 233 AU along the jet for 
TH28, RW Aur, 
and 
LkH$\alpha$321, respectively.  The only exception was in the case of the RW Aur 
blue-shifted jet lobe 
where the slit 
was placed at 0$\arcsec$.2 due to lack of line emission at 0$\arcsec$.3 \cite{Woitas02}. 
Spectra were obtained of the blue- and red-shifted lobes, using exposure times 
of 2200 and 2700 seconds respectively, for the bi-polar jets associated with 
TH28 and RW Aur. In the case of LkH$\alpha$321, only the blue-shifted lobe was 
detectable and so two spectra of this lobe, with exposure times of 2200 and 2700 seconds, 
were 
obtained and summed to increase the signal-to-noise of the faint emission lines. In 
total, this yielded five spectra, in the transverse direction at the base of 
the jets, which included H$\alpha$ and the forbidden emission lines (FELs) 
[OI]$\lambda\lambda$6300,6363, [NII]$\lambda\lambda$6548,6583, 
[SII]$\lambda\lambda$6716,6731. Data were calibrated through the standard HST 
pipeline, subtraction of the reflected stellar continuum was performed, and 
hot/dark pixels were removed. 

\section{Results}

\begin{figure*}
\begin{center}
\includegraphics[scale=0.3]{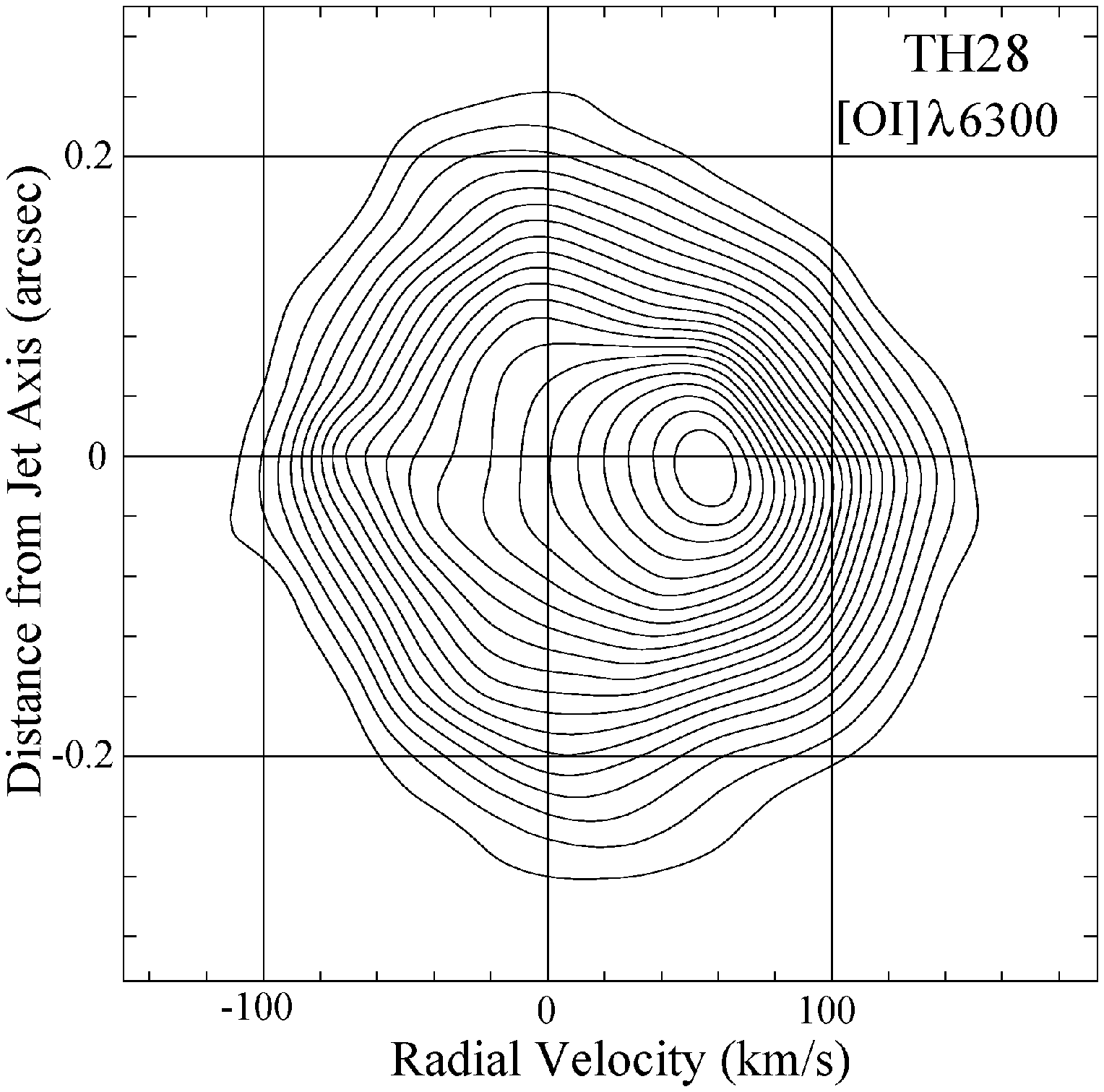}\hspace {0.27in}
\includegraphics[scale=0.3]{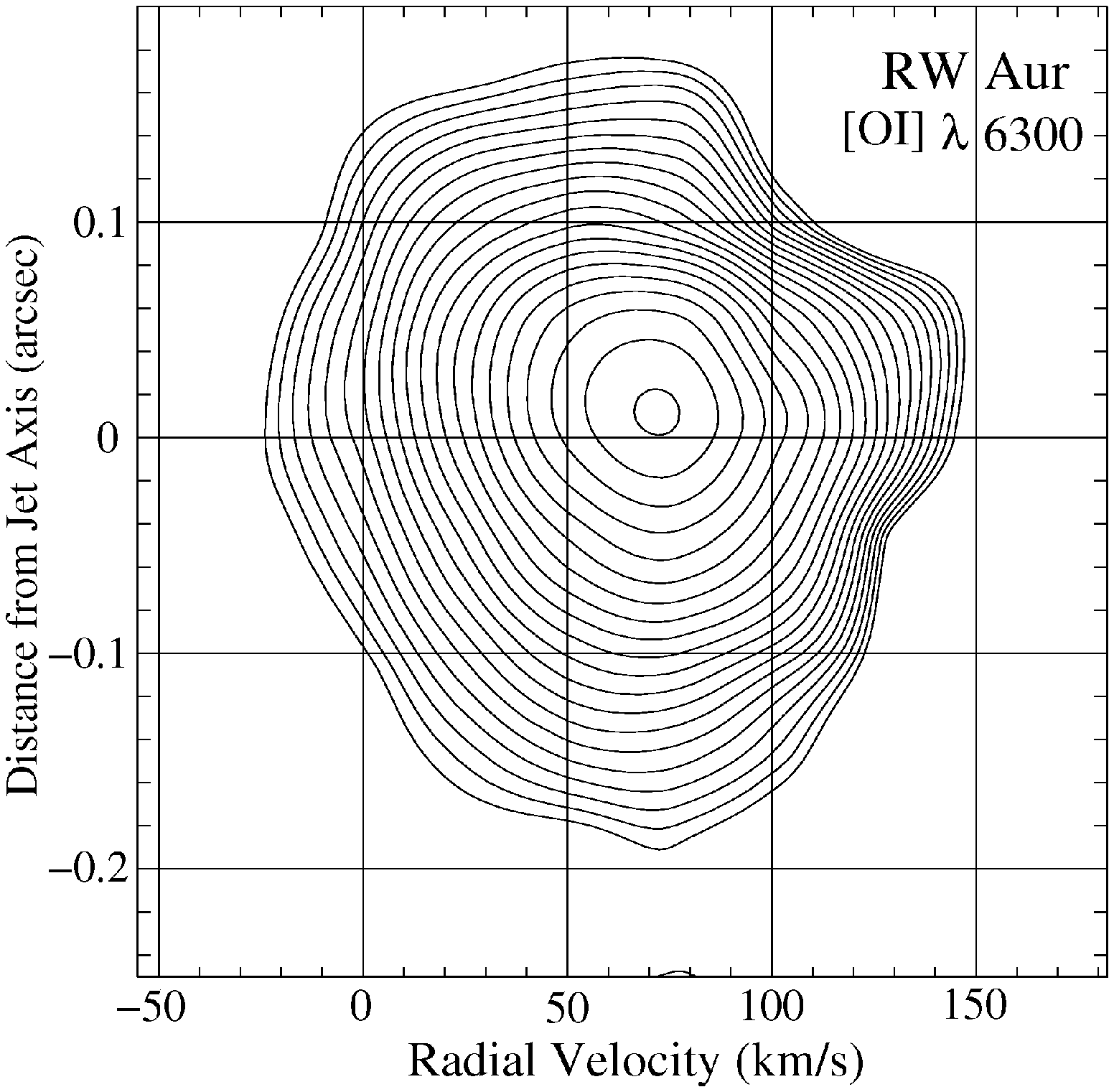}\hspace {0.27in}
\includegraphics[scale=0.3]{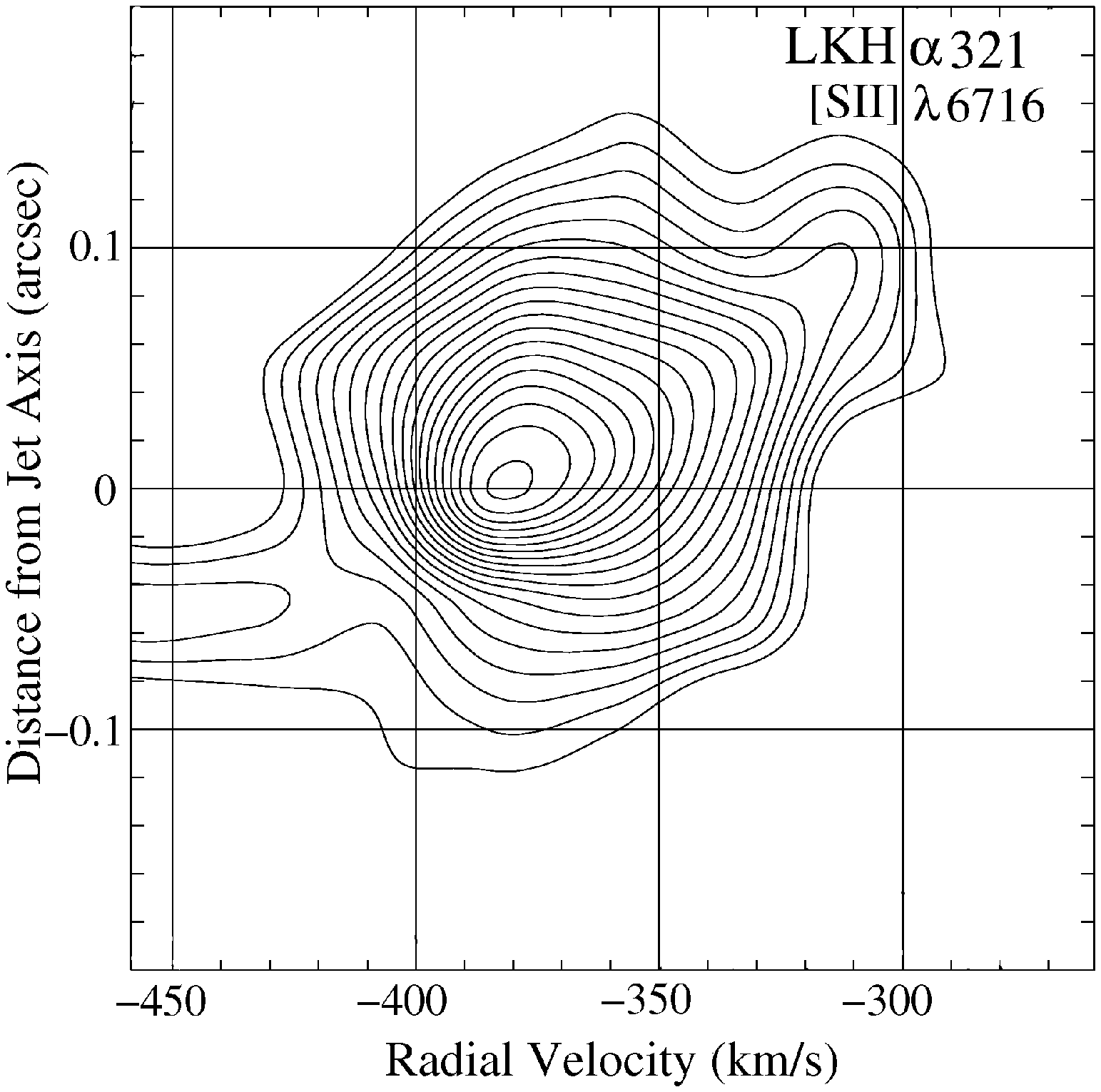}
\newline
\newline
\includegraphics[scale=0.25]{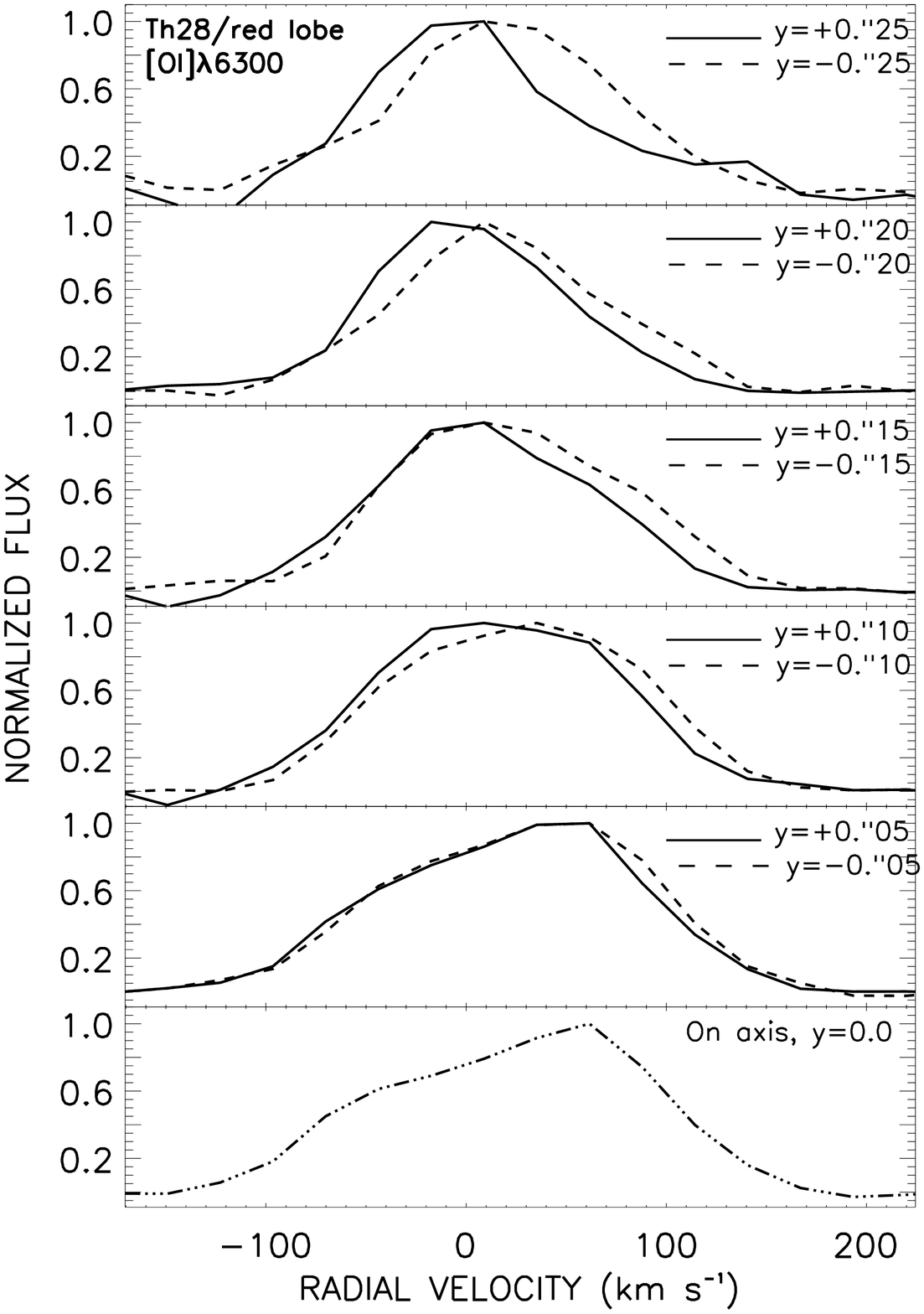} \hspace {0.25in} 
\includegraphics[scale=0.25]{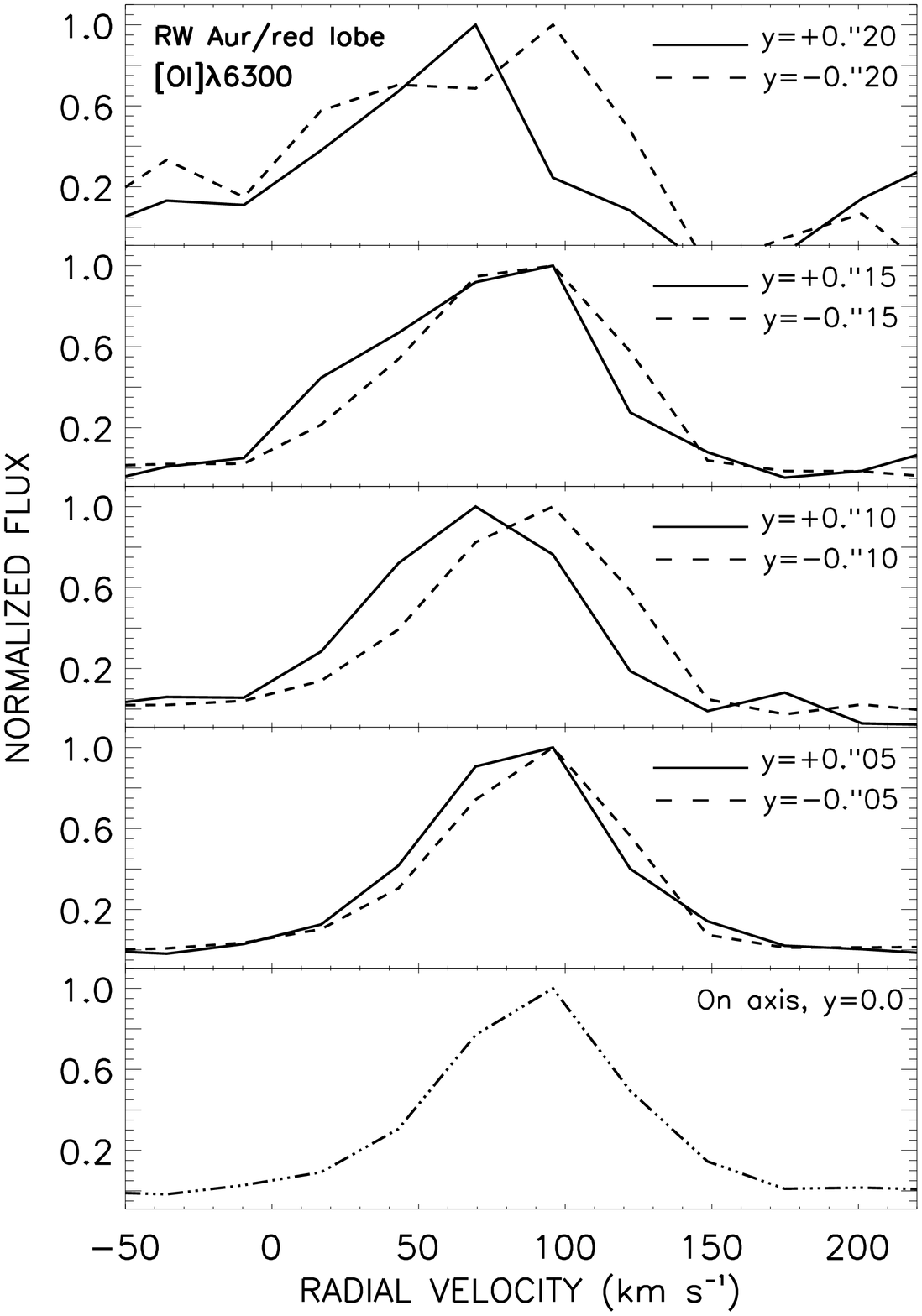} \hspace {0.25in} 
\includegraphics[scale=0.25]{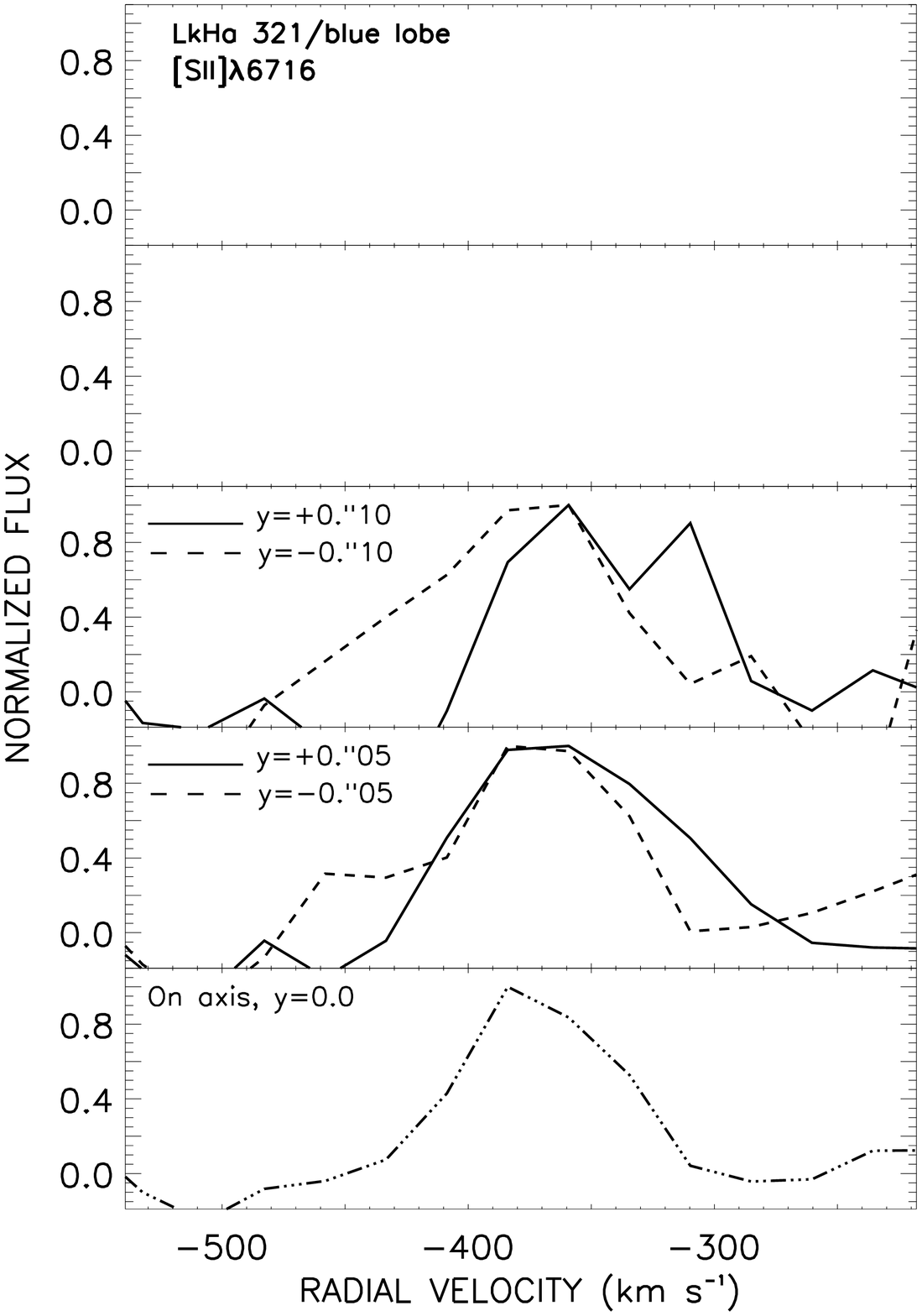}
\vspace {0.2in}
\caption{{\em Top row:} Selected position-velocity contour plots of emission lines 
from the studied jets. The skew in lower order contours is indicative of 
rotation in the outer jet channel, while the high velocity component remains 
unresolved. Contour values in units of 
erg cm$^{-2}$sec$^{-1}$ \AA$^{-1}$ arcsec$^{-2}$ in the panels are - TH28: 
from 7.0$\times$10$^{-15}$ to 6.7$\times$10$^{-14}$ with intervals 
3.0$\times$10$^{-15}$; RW Aur: from 7.2$\times$10$^{-15}$ to 
2.3$\times$10$^{-13}$, log scale interval of 2$^{1/4}$; LkH$\alpha$321: 
from 1.0$\times$10$^{-15}$ to 8.2$\times$10$^{-15}$ with intervals of 
4.0$\times$10$^{-16}$. Panels are corrected for the heliocentric velocity of the star, 
that is for +5, +23 and -7 km\,s$^{-1}$ in the three cases, respectively. 
{\em Bottom row}: Normalized intensity profiles along horizontal cuts in 
the above panels, symmetrically opposed with respect to the jet axis.}
\label{pv}
\end{center}
\end{figure*}

In order to determine whether rotation is present in the jet channel, the difference in 
velocities on 
either side of the jet axis was analysed.  Position-velocity contour plots for a sample of 
emission lines are shown in Figure\ \ref{pv}, top panels. All radial velocities are 
systemic, i.e. they are quoted with respect to the mean heliocentric velocity of the star, 
which has been measured from photospheric lines to be +5 km~s$^{-1}$ for TH28 
\cite{Graham88}, +23 km~s$^{-1}$ for RW Aur \cite{Woitas03} and -7 km~s$^{-1}$ for 
LkH$\alpha$321 (E. Whelan, priv. comm.). The lower order contours trace the outer jet 
channel where the jet is not so well collimated and where the lower velocities lie. If 
rotation is present, there will be a difference in radial velocities between the two sides 
of the jet. This difference will be evident graphically as a skew in the contours of the 
transverse position-velocity diagram. Such a skew is indeed observed in all three cases in 
the outer contour lines. This suggests the presence of rotation in at least the low 
velocity component (LVC) of the flow near the outer borders of the jet channel. The high 
velocity component (HVC), which is located much closer to the jet axis and gives rise to 
the emission peak \cite{Bacciotti00}, appears not to be spatially resolved in our spectra. 
For this reason we cannot resolve any velocity difference in the HVC between the the two 
sides of the flow, i.e. we cannot detect rotation for this velocity component.

To give a quantitative estimate of the observed velocity shifts, it was firstly necessary 
to ensure that 
we measure 
velocity offsets at equal 
distances on either side of the jet axis.
To this purpose we assumed that the peak of the (HVC) emission traces the position 
of the axis, and measured its 
distance from the nominal center of the slit, with a Gaussian fit along the 
cross-dispersion direction. We then shifted the line emission re-centering the 
HVC peak on the nominal zero arcsecond position. In all cases a small offset 
($\le$ 0.4 pixels) was required, Table\ \ref{pixel_shifts}. The offsets have different
sign and magnitude for the three targets, indicating that this is not an instrumental
systematic effect. Instead, pixel shift values 
for emission lines in opposite jet lobes are consistent with a maximum 
misalignment $\le$ 2 degrees with respect to the perpendicular of the actual jet axis
at sub-arcsecond scales.  This 
resulted in a displacement of the peak intensity of up to 0$\arcsec$.02, a
quantity which would produce a marginal asymmetry in radial velocity 
estimated to be at most 20$\%$ if uncorrected \cite{Bacciotti02}. This error 
was, however, avoided as the emission was brought on-axis prior to analysis. 

\begin{table}
{\scriptsize
\begin{center}
\begin{tabular}{lc}
\tableline \tableline
Jet lobe			&Pixel shift applied	\\ \tableline
TH28 red lobe			&+0.3672\\
TH28 blue lobe			&-0.3672\\
RW Aur red lobe			&-0.216\\
RW Aur blue lobe		&+0.1728\\
LkH$\alpha$321 blue lobe	&-0.36\\ \tableline
\end{tabular}
\end{center}
\caption{\scriptsize{Pixel shifts applied to emission lines in each jet lobe along the 
cross-dispersion 
direction, to re-center the HVC emission peak, assumed coincident with the jst axis, on 
the nominal zero 
arcsecond 
position.} \label{pixel_shifts}}}
\end{table}

\begin{figure}
\begin{center}
\includegraphics[scale=0.5]{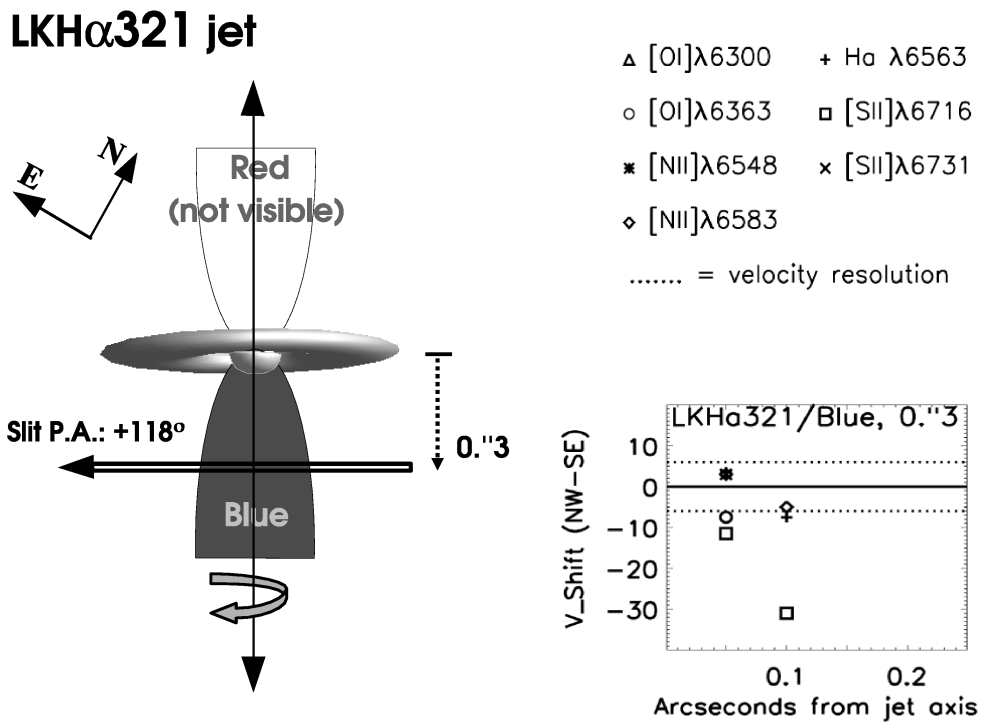}
\caption{{\em Left:} schematic drawing of the observing mode
for the blue-shifted jet from LkH$\alpha$321. The derived sense of rotation is 
illustrated by the circular arrow around the symmetry axis. 
{\em Top right:} Symbols adopted for the various lines.
{\em Bottom right:}
Summary of radial velocity asymmetries measured at the base of the blue-shifted 
lobe, at 0$\arcsec$.3 from the source, in the direction of the oriented slit, i.e.
North West  - South East (also specified 
by the label of the y-axis). 
\label{velocitydiff1}}
\end{center}
\end{figure}

\begin{figure}
\begin{center}
\includegraphics[scale=0.5]{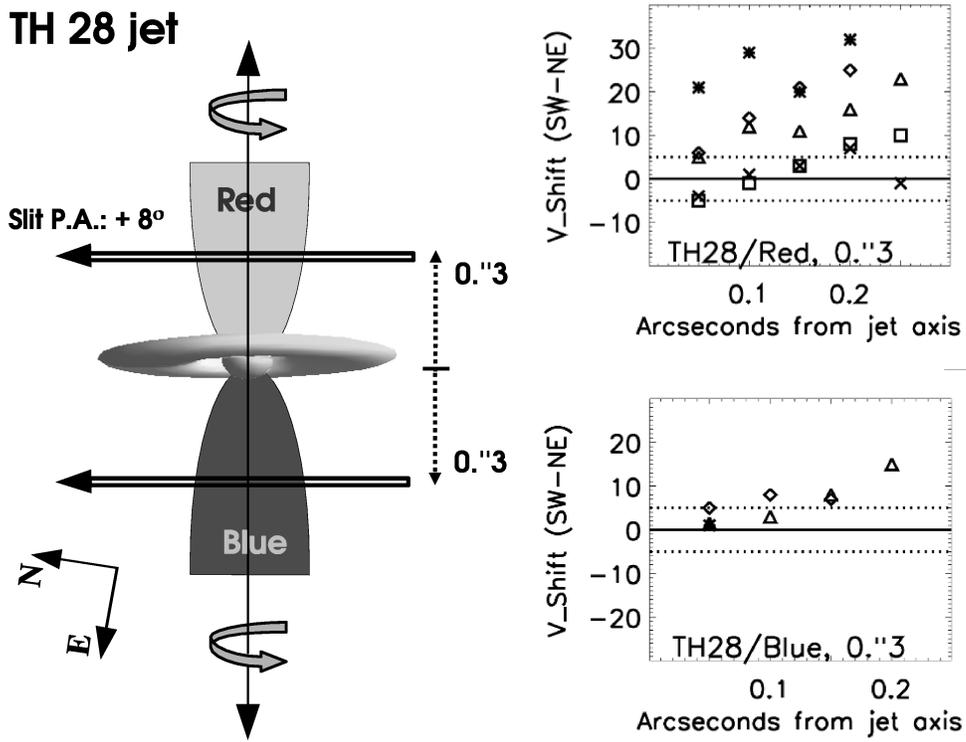}
\caption{Same as Fig.\,\ref{velocitydiff1}, 
for the bi-polar jet from TH28. Both slits were located at 
0.$''$3 from the source.
\label{velocitydiff2}}
\end{center}
\end{figure}

\begin{figure}
\begin{center}
\includegraphics[scale=0.5]{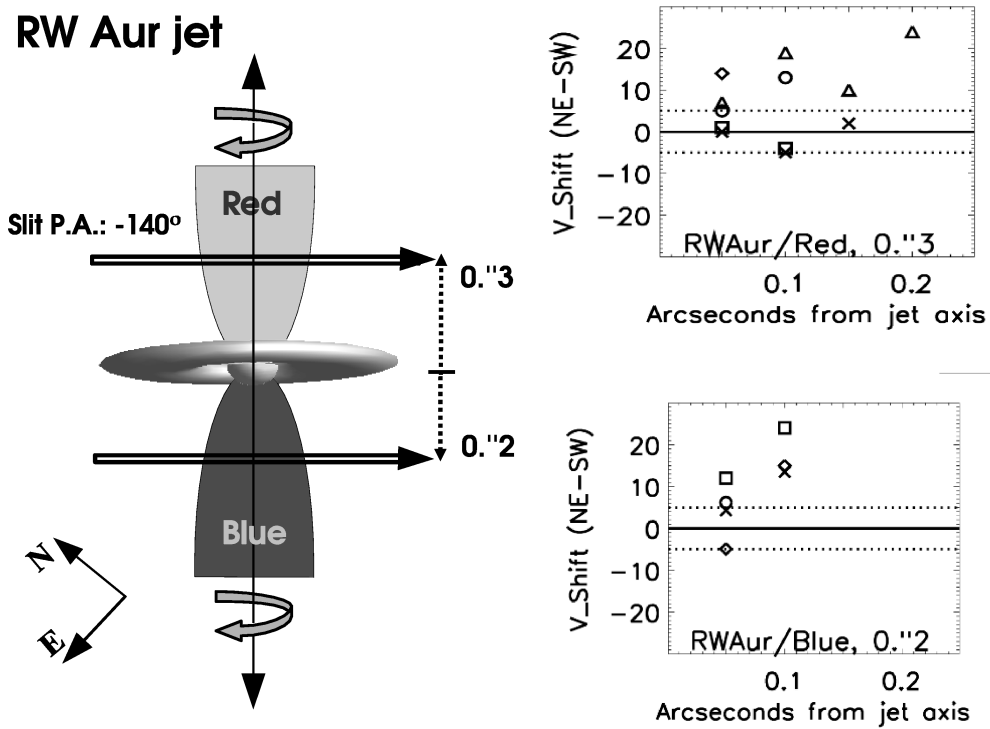}
\caption{Same as Fig.\,\ref{velocitydiff1}, 
for the bi-polar jet from RW Aur. In the  blue-shifted jet lobe 
the spectrum was taken at  0$\arcsec$.2 from the star. \label{velocitydiff3}}
\end{center}
\end{figure}

The peak intensities of pixel rows on either side of the central row were then 
compared for velocity differences, Figure\ 
\ref{pv}, bottom panels. The intensity profiles of each pair of pixel rows 
symmetric about the jet axis was plotted (e.g. $\pm$0$\arcsec$.1, or as 
indicated in each box). The single curve at the bottom in each case is the 
intensity of the central on-axis pixel row. Two methods were used in velocity
 measurements: a cross-correlation technique, which analyses the overall 
displacement of lines and is independent of the shape of the line profile; and 
a Gaussian fitting technique, which acts as a suitable check given the simple 
shape of the line profile in most cases. Specifically, each pair of 
pixel rows, mirrored in distance from the jet axis, was cross-correlated, and 
Gaussian fits for each pair of rows were also 
compared. The outcomes of the two methods are consistent, showing clear radial velocity 
differences of 10 to 25 ($\pm$ 5) km~s$^{-1}$ for opposing jet edges. 
Results are listed in Table\ \ref{radial_velocities}, in which we report 
the measured radial velocity differences in
the direction of the oriented slit, corresponding to a direction specified in the first 
column of the 
Table. 
In a small number of cases, 
flagged in the table with a $\*$ symbol, the emission had to be filtered out 
of a background which was causing velocity measurements to be artifically 
changed due to either: relatively strong HVC presence; oversubtraction of the 
background; or low signal-to-noise at crucial positions. 
The overall results are illustrated graphically in Figures \ref{velocitydiff1} to 
\ref{velocitydiff3}.

\begin{table*}
\begin{center}
\scriptsize{\begin{tabular}{llccccccc}
\tableline\tableline

Jet Lobe	&Distance		&$\Delta$V$_{rad}$ for	 	&$\Delta$V$_{rad}$ 
for 		
&$\Delta$V$_{rad}$ for		
&$\Delta$V$_{rad}$ for		&$\Delta$V$_{rad}$ for		&$\Delta$V$_{rad}$for		
&$\Delta$V$_{rad}$ 
for		
\\
		& from jet axis		&[OI]$\lambda$6300		&[OI]$\lambda$6363 		
&[NII]$\lambda$6548		
&H$\alpha$$\lambda$6563		&[NII]$\lambda$6583		&[SII]$\lambda$6716 	 	
&[SII]$\lambda$6731		
\\ 
		&(arcsec)		&(km~s$^{-1}$)				
&(km~s$^{-1}$)				
&(km~s$^{-1}$)				
&(km~s$^{-1}$)				&(km~s$^{-1}$)				
&(km~s$^{-1}$)				
&(km~s$^{-1}$)				\\ 
\tableline

TH28, red-shifted  &0.05      		&5       	&...          	&21$^{\ast}$  	& 
...      	
&6	&-5       
  	&-4		\\
(SW - NE)	&0.1	         	&12       	&...          	&30$^{\ast}$	
&...       	
&14	&-1       
  	&1		\\
        	&0.15      	  	&11       	&...          	&30$^{\ast}$  	
&...       	
&21	&3        
  	&3		\\
        	&0.2       	  	&16       	&...          	&32$^{\ast}$ 	
&...       	
&25	&8        
  	&7		\\
        	&0.25       	 	&23       	&...          	&...         	
&...       	
&...	&10    
  	&-1		\\
		&			&		&		&		&		
&	&	
	& 		\\  
TH28, blue-shifted &0.05 		&2$^{\ast}$  	&...            &1		
&...            
&5	&...  
          &...            \\  
(SW - NE)	&0.1       	 	&3$^{\ast}$	& ...           &...            
&...            
&8	&...  
          &...            \\              
        	&0.15      	 	&8       	&...            &...            
&...            
&7	&...  
          &...            \\              
        	&0.2        		&15       	&...            &...            
&...            
&...	&...  
          &...            \\          
		&			&		&		&		&		
&	&	
	& 		\\ 
RW Aur, red-shifted &0.05  		&7         	&5        	&...        	
&...            
&14     &1      
  	&-1		\\
(NE - SW)	&0.1       	  	&20        	&14        	&...        	
&...       	
&...    &-4     
  	&-5		\\
        	&0.15      	  	&10        	&...        	&...        	
&...       	
&...    &...  
          &2		\\
        	&0.2        	 	&24        	&...        	&...        	
&...       	
&...    &...  
          &... 		\\
		&			&		&		&		&		
&	&	
	& 		\\ 
RW Aur, blue-shifted &0.05 		&...            &6$^{\ast}$ 	&...        	
&...            
&-5$^{\ast}$ 	
&12     &4$^{\ast}$	\\
(NE - SW)	&0.1       	 	&...            &        	&...        	
&...            
&15$^{\ast}$  	
&25     &14$^{\ast}$	\\
		&			&		&		&		&		
&		
&	& 		\\ 
LkH$\alpha$321, blue-shifted &0.05	&...		&-8$^{\ast}$	&...        	
&4$^{\ast}$	
&3$^{\ast}$	
&-12$^{\ast}$	&3$^{\ast}$	\\
(NW - SE)	&0.1     		&...		&...		&...		
&-7$^{\ast}$	
&-5$^{\ast}$	
&-31$^{\ast}$	&...		\\ \tableline

\end{tabular}}
\end{center}
\caption{\scriptsize{Radial velocity differences, $\Delta$V$_{rad}$, across the jet 
0$''$.3 from the 
source
(0$''$.2 for the RW Aur blue lobe), 
measured using both the cross-correlation technique and single Gaussian fitting, along the 
direction 
specified in the first column. The symbol  $^{\ast}$ marks data points which have 
been filtered out of a low S/N environment, 
or from a strong spurious contribution from the HVC emission wings. 
Where dots appear in the table, the emission was either shifted off the 
CCD, or was too faint, small or scattered to decipher. 
The accuracy reached with the data analysis is approximately $\pm$5 
km~s$^{-1}$.}\label{radial_velocities}}
\end{table*}

\begin{figure*}
\begin{center}
\includegraphics[scale=0.65]{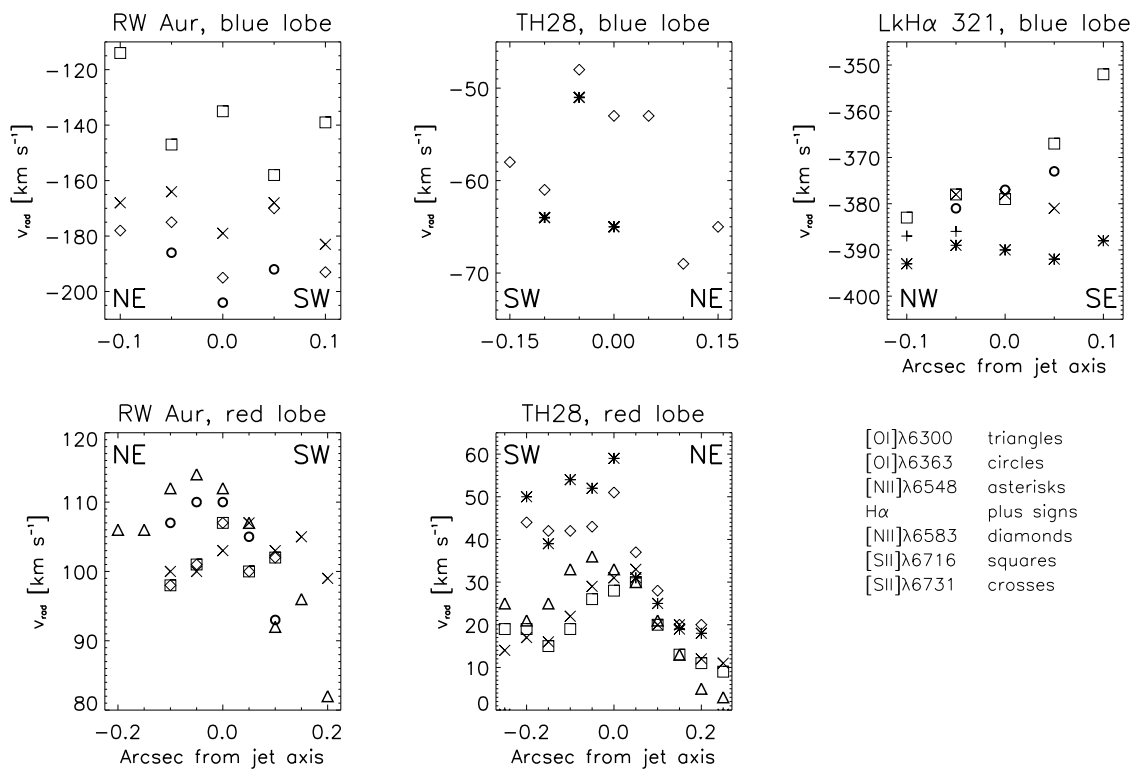}
\caption{Radial velocity profile across each jet in various emission lines. 
The TH28 blue-shifted 
[OI] $\lambda$6300 line is not included as the profile of this emission line is very wide
and  did not 
allow measurement of velocity peaks with Gaussian fitting. In this case the rotational 
velocities were derived from cross-correlation routines alone, resulting only 
in velocity difference measurements. 
\label{transvel}}
\end{center}
\end{figure*}

Finally, the radial velocity profile across each jet is shown in Figure\ \ref{transvel}. 
In the 
red-shifted lobes of TH28 and RW Aur, the on-axis radial velocity is clearly the highest 
and the value 
reduces as the edges of the jet are approached. The  blue-shifted lobes are not so clear 
as the emission 
is fainter and often more scattered. Data reduction in these cases often required 
isolation of the lower 
velocity component from the HVC (as marked in Table\ \ref{radial_velocities}) to identify 
elements of 
rotation. However radial velocities in this figure {\em include} the HVC, and so rotation 
is not 
apparent in all cases. 

From the results of this spectral analysis, combined with the inclination angles, we find 
poloidal 
velocities for the RW Aur jet of 144 - 227 km~s$^{-1}$ in the red lobe and 245 - 288 
km~s$^{-1}$ in the 
blue lobe - this velocity asymmetry is well known from previous observations, 
\cite{Woitas02}; for TH28 
jet of 115 - 288 km~s$^{-1}$ in the red lobe and 230 - 374 km~s$^{-1}$ in the blue lobe - 
an asymmetry 
which was also previously recorded, \cite{Graham88}; and for LkH$\alpha$321 of 540 - 550 
km~s$^{-1}$. 
Toroidal velocities derived from the outer positional radial velocity shifts, being less 
affected by 
projection effects (see Section 4), are in the ranges of: 7 to 17 km~s$^{-1}$ for both 
lobes of RW Aur; 
5 to 13 km~s$^{-1}$ for the red lobe of TH28; 4 to 8 km~s$^{-1}$ for the blue lobe of 
TH28; and 4 to 9 
km~s$^{-1}$ for the blue lobe of LkH$\alpha$321. (Note that, in the ranges given above, 
the higher poloidal velocities lie at distances closest to the rotation axis, and these 
correspond to the lower toroidal and radial velocities.) 

\section{Discussion}

Assuming that emission from the jet is axially symmetric, we interpret our 
findings of velocity differences between the two sides of the jet flow, Table\ 
\ref{radial_velocities} 
and Figures\ 
\ref{velocitydiff1} to \ref{velocitydiff3}, as evidence for rotation at the 
base of the jet. Most importantly, for TH28 and RW Aur, the red and blue jet 
lobes were found to rotate in the same direction. This implies that the 
helicity in the red and blue lobes (i.e. the handedness of toroidal with respect to the 
poloidal 
velocity) is 
opposite in opposite directions, a result predicted by MHD 
models where the ambient field is wrapped around due to disk rotation. 

Before looking at each target individually, there are a few general comments 
to be made about the results in Table\ \ref{radial_velocities}. Firstly, the velocity 
difference 
measurements close to the jet axis, i.e. at the 0$\arcsec$.05 position, are noticeably 
smaller than points further away, and where the emission is detectable at 0$\arcsec$.2 and 
beyond, velocity differences are higher than at intermediate distances. 
This effect may at first appear in contradiction with the notion that the central 
portions of the jet should rotate faster. A detailed comparison with disk-wind 
model predictions (\citealp{Pesenti03}, and \citealp{Dougados03}) shows, however, that the 
apparent decrease of the observed velocity difference toward the jet axis is likely to be 
due to 
projection and beam smearing effects. Since the emission is optically thin we see, along 
the line of sight, the sum of the contributions emitted from regions in the jet that 
rotate with different toroidal velocities. This causes  a reduction of the observed 
velocity shift. Such an effect is more important for regions closer to the axis, while the 
values measured at the outer jet borders are less contaminated and so are in better 
agreement with 
theoretically 
predicted toroidal velocities. In other words this effect does not reflect a true 
kinematic feature,
but is expected on the basis of MHD acceleration models when combined with 
our observational mode. Secondly, it should be noted that the size of velocity differences 
in 
different emission lines does not represent scattering around an average value 
but rather is due to the fact that emission has its origin at different 
positions along the line of sight. And lastly, the [NII] lines show higher 
velocity differences than other emission lines, a result that illustrates how 
they trace the central more collimated higher velocity region of the flow 
(see \citealp{Bacciotti00}, \citealp{Martin03}, \citealp{Pesenti03}). 

The results for the TH28 red-shifted jet lobe are clearest. Velocity 
differences alomg the SW - NE direction are positive, with a few 
exceptions mainly in the [SII] lines. Also, the data close to the jet axis 
are not  well resolved as explained above. 
The stronger [OI] and [NII] lines have values of 5 and 6 km~s$^{-1}$ at 
0$\arcsec$.05 from the axis 
compared to 10 to 20 km~s$^{-1}$ further from the axis, 
while the outer jet channel seems to have higher radial velocity 
differences of about 24 km~s$^{-1}$. Values in the blue-shifted 
jet lobe are less clear, but [NII] and [OI] emission give positive differences 
consistent with the red-shifted 
lobe. All other usable data points fall within the error bars about zero. Globally, we 
find evidence 
that both lobes 
of 
the jet rotate in a clockwise 
direction, looking down the blue-shifted jet lobe towards the source, with a measured
radial velocity difference  of 10 to 25 km~s$^{-1}$. 

The RW Aur jet also shows clear indications of rotational velocities. 
Exceptions include the [SII]$\lambda$6731 values in the red-shifted lobe. Also points 
close to the jet 
axis show 
smaller radial velocity differences due to strong unresolved HVC emission. However, the 
[OI] lines give 
clear results 
with higher radial velocity difference evident at 0$\arcsec$.2 from the jet axis. The 
blue-shifted lobe 
is less 
definite, but nevertheless 
velocity differences outside the error bars are positive in line with the 
red-shifted lobe. Overall, results show an anti-clockwise rotation looking 
down the blue-shifted jet lobe towards the source, again with radial velocity differences 
of 10 to 
25 km~s$^{-1}$. These findings are consistent in magnitude and direction with 
results of similar research on the RW Aur jet \cite{Woitas03} in which rotational 
velocities in the same 
direction of 
10 to 20 km~s$^{-1}$ have been observed, in the 
form of velocity differences between the borders of the flow. For this study HST/STIS was 
also used, but 
the spectra 
were taken in a set of positions across the jet with the slit direction parallel to the 
jet axis.  

In the case of LkH$\alpha$321, which is located at 550 pc (more than three times the 
distance of the 
other targets) the emission lines were very faint despite having combined two spectra to 
increase the signal-to-noise ratio. Nevertheless, velocity differences in the range of -5 
to -30 km~s$^{-1}$ have been measured, although evidence of rotation is weaker in this 
case. The sense of rotation for LkH$\alpha$321 is measured as anti-clockwise looking down 
the blue-shifted lobe towards the star.  In this jet we are able to measure rotation at 
higher velocities because of a number of factors: LkH$\alpha$321 is farther away meaning 
we are looking further along the jet (233 AU from the source, projected distance) to a 
point where it has widened and so we can resolve higher velocities; a large inclination 
angle (which appears to be the case from spectro-astrometric measurements \cite{Whelan03}) 
would mean that the poloidal and radial velocities approach each other; and finally this 
is a larger mass T~Tauri star, since it has spectral type G1 \cite{Chavarria81}, implying 
higher velocities in the outer resolvable regions of the jet. 

It is reassuring that we obtain {\em negative} shifts in the case of LkH$\alpha$321, as 
opposed to the 
positive shifts observed for the other two sources because, as we are comparing the {\em 
same} rows on 
the CCD detector in all cases, it means we are not measuring an instrumental effect. 
Conceiveably, a slight misalignment of the slit with respect to the transverse direction 
of the jet (i.e inaccurate position angle used in the pointing of the 
instrument) may produce an effect similar to rotation. In this case, the position of the 
real jet axis will then be shifted with 
respect to the nominal zero arcsec position. We have shifted the spectral image back to 
the zero arcsec 
position as previously discussed (Section 3), but the angle subtended remains a problem. 
It could 
produce a rotation signature even in a non-rotating jet, since the HVC and LVC are at 
different spatial 
locations on the CCD with respect to the zero arcsec row of pixels, implying that we are 
not probing 
symmetric regions of the jet with respect to its real axis. Given such a misalignment has 
occurred, the 
extent of the contamination does not have, however, a dramatic affect on our results, as 
can be seen 
from the fact that the pixel shift requirements for TH28 and LKH$\alpha$321 (Table\ 
\ref{pixel_shifts}) 
are in the same direction but the sense of rotation of their jets is opposite. For RW Aur, 
the same PA was used 
here for slit positioning as in a previous study \cite{Woitas03}, where it was found that 
the magnitude 
of the false rotation signature contamination due to incorrect PA was, at most, 1 - 5 
km~s$^{-1}$. 
However, the sense of the false signature is in fact opposite in direction to that of the 
jet's 
rotation, and so the values we have measured are actually lower limits. Apart from this, 
the only other 
obvious effect which could produce a contour skew mimicking rotation is asymmetrical 
interaction with 
the local environment on either side of the propagating jet, e.g. asymmetrical mass 
entrainment leading to asymmetrical poloidal velocities. However 
such mimicking is unlikely because (apart from the fact that asymmetrical entrainment 
should also produce enhanced emission at one border of the flow, which is not seen in our 
spectra) we see the same asymmetry in both the red and blue shifted jet lobes where 
present, and the magnitudes of the deduced toroidal velocity differences are in the range 
predicted by theory (as discussed below). 

Overall, our observations are in line with the observations of the jet from the T~Tauri 
star DG Tau \cite{Bacciotti02}. In that case, by using simple and general relationships 
governing the physics of magnetically launched disk~winds together with an observationally 
based estimate of the ratio, $R$, between the mass flux in the jet and the mass flux 
accreted through the disk ($R$ $\sim$ 0.1), it was demonstrated that the observed velocity 
differences were in the expected range, \cite{Bacciotti02, Anderson03, Dougados03}. These 
values compare well with our results, which therefore support the magneto-centrifugal 
scenario. 
Furthermore, our toroidal and poloidal velocities have the same ratio as theoretical 
predictions \cite{Vlahakis00}, and we can use these velocities, $v_{\phi,\infty}$ and 
$v_{p,\infty}$ measured at a distance, $\varpi_{\infty}$, from the rotation axis, to 
obtain values for the wind-launch region in 
terms of distance from the rotation axis along the disk plane, $\varpi_{0}$. 
Using Equation\ \ref{foot_point} \cite{Anderson03},  
\begin{eqnarray}
\varpi_{0} & \approx & 0.7AU 
\left(\frac{\varpi_{\infty}}{10AU}\right)^{2/3}\ 
\left(\frac{v_{\phi,\infty}}{10kms^{-1}}\right)^{2/3}\ \nonumber \\ & & \times 
\left(\frac{v_{p,\infty}}{100 
kms^{-1}}\right)^{-4/3}\ \left(\frac{M_{\star}}{1M_{\odot}}\right)^{1/3} 
\label{foot_point}
\end{eqnarray}
and assuming all three sources are of mass M$_{\star}$ $\sim$ 1M$_{\odot}$, which is a 
reasonable 
approximation given 
the weak dependence on M$_{\star}$, we obtain values for $\varpi_{0}$ as shown in Table\ 
\ref{launch_point}. We have 
chosen measurements at 0$\arcsec$.1 and 0$\arcsec$.2 from the jet axis as limits of a 
suitable range, being less 
contaminated by 
projection effects. Where emission was faint, values corresponding to 0".05 were used, but 
it should be 
noted that 
these are less accurate. Also, the values at 0$\arcsec$.1 are less precise than those at 
0$\arcsec$.2, when both are present, since increased projection effects close to the jet 
axis tend to reduce the line of sight averaged v$_{\phi,\infty}$, and hence the resulting 
value of $\varpi_{0}$ (Pesenti et al 2003). 
Considering the red-shifted jet-lobes which both have very strong signal-to-noise in [OI] 
and [NII] 
emission lines, 
we calculate a wind-launch region spanning 0.3 to 1.6 AU, in keeping with previous 
estimates for DG Tau 
of $\sim$ 
1.8~AU \cite{Bacciotti02} and $\sim$ 0.3 to 4~AU \cite{Anderson03}. The observational 
evidence presented here supports the idea that disk
winds are launched, via the magneto-centrifugal mechanism (e.g. \citealp{Konigl00}), at radii within a few AU of the star. Determination of
the most appropriate model however will have to await higher spatial and
spectral resolution observations in the future. 


\begin{table}
\begin{center}
\scriptsize{
\begin{tabular}{llccccc}
\tableline \tableline
Star	&Jet lobe &$\varpi_{\infty}$	&$\varpi_{\infty}$	&$\triangle v_{radial}$	
&$v_{radial}$	
&$\varpi_{0}$ 	
\\
	&	&arcsec			&AU			&km~s$^{-1}$		
&km~s$^{-1}$	
&AU		
\\ \tableline

TH28 	&Red-shifted	&0.1 - 0.2		&17 - 34		&10 - 25		
&30 - 20	
&0.3 
- 1.6	\\ 
	&Blue-shifted	&0.1 - 0.2		&17 - 34		&8 - 15			
&50 - 40	
&0.1 
- 0.4	\\

RW Aur	&Red-shifted	&0.1 - 0.2		&14 - 28		&10 - 25		
&105 - 
100	&0.4 
- 1.3	\\ 
	&Blue-shifted	&0.05 - 0.1		&7 - 14			&10 - 25		
&180 - 
170	
&0.1$^{\ast}$ - 0.4	\\ 

LkH$\alpha$321 &Blue-shifted &0.05 - 0.1	&27.5 - 55		&5 - 12			
&390 - 
380	
&0.1$^{\ast}$ - 0.2	\\ \tableline

\end{tabular}
}
\end{center}
\caption{The range for the launch point of the disk wind, $\varpi_{0}$, for our five 
targets, calculated 
using the 
method described in Anderson et al (2003). $^{\ast}$Where emission is faint, values for 
the 0".05 
position were used, 
but these should be considered less accurate due to resolution constraints. 
\label{launch_point}}
\end{table}

\section{Conclusions}

The jets from the three young stars observed, TH28, RW Aur and LkH$\alpha$321, show 
distinct 
and systematic radial velocity asymmetries in opposing positions with respect to the jet 
axis, at 0$''$.2 - 0$''$.3 from the source. Although the on-axis higher velocity component 
of the jet remains unresolved, radial velocity differences in the lower velocity component 
located in the outer jet channel are found to be on the order of 10 to 25 ($\pm$5) 
km~s$^{-1}$. For the bi-polar jets from TH28 and RW Aur, the velocity differences have the 
same sign in both 
lobes. We interpret these radial velocity asymmetries as rotation signatures in the region 
where the jet has been collimated but has not yet manifestly interacted with the 
environment. Therefore the sense of rotation of the jets, looking down the blue-shifted 
lobe towards the star, is clockwise for TH28, and anti-clockwise for RW Aur and 
LkH$\alpha$321.

Our findings are reinforced in a number of ways: the velocity differences are of the same 
magnitude as those measured in the similar DG Tau jet of 5 to 10 km~s$^{-1}$ 
\cite{Bacciotti02}, which was shown to be in agreement with the predictions of MHD 
disk-wind models \cite{Bacciotti02, Anderson03, Dougados03, Pesenti03}; they are in line 
with 
similar research on the RW Aur jet \cite{Woitas03} which yields rotational velocities of 
10 to 20 km~s$^{-1}$, with the same sense of rotation; and finally, they lead to values 
for the distance of the LVC footpoint from the central axis of $\approx$ 0.5 - 2 AU, 
consistent with the models of magneto-centrifugal launching \cite{Anderson03}.  

\vspace {0.2in}
{\bf Acknowledgements} 
\vspace {0.1in}
\newline
We wish to thank Marcello Felli, Catherine Dougados, Emma Whelan and Jonathan Ferreira for 
useful comments and suggestions. D.C. and T.P.R. would like to acknowledge 
support for their research from Enterprise Ireland and J. E. and J. W. 
likewise wish to acknowledge support from the Deutsches Zentrum f\"ur Luft- und Raumfahrt 
under grant 
number 50 OR 
0009. We would also like to thank the anonymous referee for useful comments.


\clearpage


\begin{thebibliography}{}
{ 

\bibitem[Anderson et al. 2003]{Anderson03}
Anderson, J. M., Li, Z.-Y., Krasnopolsky, R. \& Blandford, R., 2003, \apj, 590, L107 

\bibitem[Bacciotti \& Eisl\"offel 1999]{BE99}
Bacciotti, F. \& Eisl\"offel, J., 1999, A$\&$A, 342, 717

\bibitem[Bacciotti et al. 2000]{Bacciotti00}
Bacciotti, F., Mundt, R., Ray, T. P., Eisl\"offel, J., Solf, J. \& Camezind, M. 2000, 
\apj, 537, L49

\bibitem[Bacciotti et al. 2002]{Bacciotti02}
Bacciotti, F., Ray, T. P., Mundt, R., Eisl\"offel, J. \& J., Solf, 2002, \apj, 576, 222

\bibitem[Chavarr\'{i}a-K. et al. 1981]{Chavarria81}
Chavarr\'{i}a-K., C., \& de Lara, E., RMxAA, 1981, 6, 159C

\bibitem[Davis et al. 2000]{Davis00}
Davis, C. J., Berndsen, A., Smith M. D., Chrysostomou, A. \& Hobson, J., 2000, MNRAS, 314, 
241 

\bibitem[Dougados et al. 2003]{Dougados03}
Dougados, C., Cabrit, S., Ferreira, J., Pesenti, N., Garcia, P. \& O'Brien, D., 
in proc of ``Magnetic Fields and Star Formation, theory versus observations'', 
2003, AP\&SS in press  

\bibitem[Ferreira 1997]{Ferreira97}
Ferreira, J., 1997, A\&A, 319, 340

\bibitem[Graham et al. 1988]{Graham88}
Graham, J. A. \& Heyer, M. H., 1988, PASP, 100, 1529

\bibitem[K\"onigl \& Pudritz 2000]{Konigl00}
K\"onigl, A. \& Pudritz, R., 2000, in Protostars and Planets IV, V. Mannings, A. P. Boss, 
$\&$ S. S. 
Russell 
(Tuscon: Univ. Arizona Press), 759

\bibitem[Krautter 1986]{Krautter86}
Krautter, J., 1986, A$\&$A, 161, 195

\bibitem[Kwan \& Tademaru 1988]{Kwan88}
Kwan, J. \& Tademaru, E., 1988, \aj 332L, 41K

\bibitem[L\'{o}pez-Mart\'{i}n et al. 2003]{Martin03}
L\'{o}pez-Mart\'{i}n, L., Cabrit, S. \& Dougados, C., 2003, A$\&$A, 405, L1

\bibitem[Mundt \& Eisl\"offel 1998]{Mundt98}
Mundt, R. \& Eisl\"offel, J., 1998, \aj 116, 860

\bibitem[Pesenti et al. 2003]{Pesenti03}
Pesenti, N., Dougados, S., Cabrit, S., Ferreira, J., Casse, F., Garcia, P. \& O'Brien, D., 
2003, A$\&$A, 
submitted

\bibitem[Shu et al. 2000]{Shu00}
Shu, F. H., Najita, J. R., Shang, H. \& Li, Z.-Y., 2000, in Protostars and Planets IV, V. 
Mannings, A. 
P. Boss, 
$\&$ S. S. Russell (Tuscon: Univ. Arizona Press), 789

\bibitem[Testi et al. 2002]{Testi02}
Testi, L., Bacciotti, F., Sargent, A. I., Ray, T. P. \& Eisl\"offel, J., 2002, A$\&$A, 
394, 31

\bibitem[Vlahakis et al. 2000]{Vlahakis00}
Vlahakis, N., Tsinganos, K., Sauty, C. \& Trussoni, E., 2000, MNRAS, 318, 417

\bibitem[Whelan et al. 2003]{Whelan03}
Whelan, E. T., Ray, T. P., Davis, C. J. \&  2003, A$\&$A, submitted

\bibitem[Woitas et al. 2002]{Woitas02}
Woitas, J., Ray, T. P., Bacciotti, F., Davis, C. J. \& Eisl\"offel, J., 2002, \apj, 580, 
336

\bibitem[Woitas et al. 2003]{Woitas03}
Woitas, J., Bacciotti, F., Ray, T. P., Marconi, A., Coffey, D. \& Eisl\"offel, J., 2003, 
A$\&$A, 
submitted

}
\end{thebibliography}
\end{document}